\documentclass{osa-article}

\journal{oe}

\usepackage{lipsum} 
\usepackage{upgreek}
\usepackage{gensymb}

\usepackage{soul} 

\begin{document}

\title{A dual color, frequency, pulse duration and shape agile laser system for particle spectroscopy and manipulation}

\author{Junhwi Bak,\authormark{1}, Robert Randolph\authormark{1}, and Alexandros Gerakis,\authormark{1,2,*}}

\address{\authormark{1}Department of Aerospace Engineering, Texas A\&M University, College Station, TX, 77843, USA}
\address{\authormark{2}Luxembourg Institute of Science \& Technology, 4422 Belvaux, Luxembourg}

\email{\authormark{*}alexandros.gerakis@list.lu} 



\begin{abstract}
A dual color, frequency and pulse duration agile laser system, capable of delivering laser pulses in arbitrary temporal profiles with $\sim1$ ns to $\sim1$ $\upmu$s pulse duration, chirping rates of $\sim27$ MHz/ns with an achievable chirping range of several GHz across the pulse duration, and with energies ranging from a few nJ to hundreds of mJ per pulse has been developed. The flexibility and capability of this laser system provide a wide range of laser parameters that can be exploited to optimize operational conditions in various experiments ranging from laser diagnostics to spectroscopy and optical manipulation of matter. The developed system is successfully demonstrated to obtain coherent Rayleigh-Brillouin scattering (CRBS) in both single and dual color configuration, in an effort to expand the non-intrusive accessibility towards lower pressure regime for neutral gas and plasma diagnostics.
\end{abstract}

\section{Introduction}
Since their initial demonstration by Maiman~\cite{maiman1960stimulated}, lasers have gradually become perhaps the most important tool for neutral gas, plasma and nanoparticle diagnostics, both stagnant and in a flow. Exhibiting the significant advantages of performing fast, non-intrusive and non-perturbative measurements, the ultimate goal of laser diagnostics is to replace the use of mechanical probes that are still the state of the art for relevant measurements, such as Pitot tubes and hot-wires for neutral gas flows~\cite{shinder2015nist,4198449,RUSSO201167,gatski2013compressibility,morris2012measurement,JENSEN2004}, Langmuir and emissive probes in the case of plasma flows~\cite{10.1063/1.5016229,cherrington1982use} and mechanical extractors in the case of nanoparticles~\cite{FANG2016273}. 

Mechanical probes are inherently intrusive to the system they measure, and they exhibit generally slow response times while sometimes they do not survive the environment which they are called to measure. To overcome these limitations, a plethora of laser based diagnostic techniques have been developed and demonstrated over the years to help measure thermodynamic quantities of interest in all states of matter. Those designed to perform measurements on neutral gases include, to name but a few, Spontaneous Rayleigh Scattering (SRayS)~\cite{Miles_2001,Kampmann:93}, Filtered Rayleigh Scattering (FRS)~\cite{doi:10.1366/000370208784909526,gustavsson2005filtered,Doll:17,Hoffman:96,forkey1993time,KRISHNA2021329}, Coherent Rayleigh-Brillouin Scattering (CRBS)~\cite{Pan2,Gerakis2, doi:10.1063/1.4959778}, Spontaneous Raman Scattering (SRS)~\cite{Karpetis:96,doi:10.1002/jrs.1965}, Coherent Anti-Stokes Raman Scattering (CARS)~\cite{Tolles:s,zheltikov,Lempert_2014,doi:10.1002/jrs.1489,Lempert:11,doi:10.1002/jrs.5339,doi:10.1063/1.3483871}, Laser Induced Fluorescence (LIF) and Planar LIF (PLIF)~\cite{hanson1990planar,Lee:93,crimaldi2008planar,jiang2012no,doi:10.2514/2.1939}. 
In addition to diagnostics, advanced neutral and plasma manipulation concepts such as laser acceleration of neutral species~\cite{maher2012laser}, optical deceleration of atomic~\cite{https://doi.org/10.48550/arxiv.2206.07034} or molecular species~\cite{PhysRevLett.93.243004,Yang_2018} to even antihydrogen transport~\cite{Madsen_2021}, as well as dual laser pulse ignition schemes~\cite{doi:10.1063/1.5043295,f96eb5b913034b98b912310b23394b33}, can also benefit from the use of advanced laser sources. 

Although the techniques mentioned above have already been demonstrated using mostly standard, commercial laser sources, their efficiency has often been subject to limitations posed by the performance specifications and capabilities of the laser sources used in the respective studies. Specifications such as pulse duration, energy per pulse, temporal shape of the laser pulses and, in some cases, achievable chirp of the pulses are mainly fixed parameters, essentially limiting the exploration and determination of the optimum operating conditions per use case. 

To enable better efficiency and enhancement of the operational parameter space in each of these areas, a novel laser system has been developed, applicable (non exclusively) to all of the aforementioned regimes and techniques, and is presented here. This system is capable of achieving (i) variable pulse durations ranging from $\sim1$ ns to $\sim1$ $\upmu$s with (ii) arbitrary temporal profiles at (iii) a dual wavelength configuration, while (iv) exhibiting faster chirping rates to what has been reported in the relevant literature, up to a full chirping range of $\sim10$ GHz. Additionally, the system presented here (v) can be easily scalable, able to deliver pulse energies ranging from a few nanoJoules per pulse up to a few Joules per pulse, with no physical limitation on the achievable upper limit. A laser system like the one presented here can bring substantial advances to the fields discussed previously, by not only performing more optimally versus current laser designs but can also, and more importantly, lead to novel modi operandi of the techniques where it is being used.

\section{A dual-color, frequency and pulse duration agile laser system}
\begin{figure}[h!]
\centering
\includegraphics[width=\textwidth]{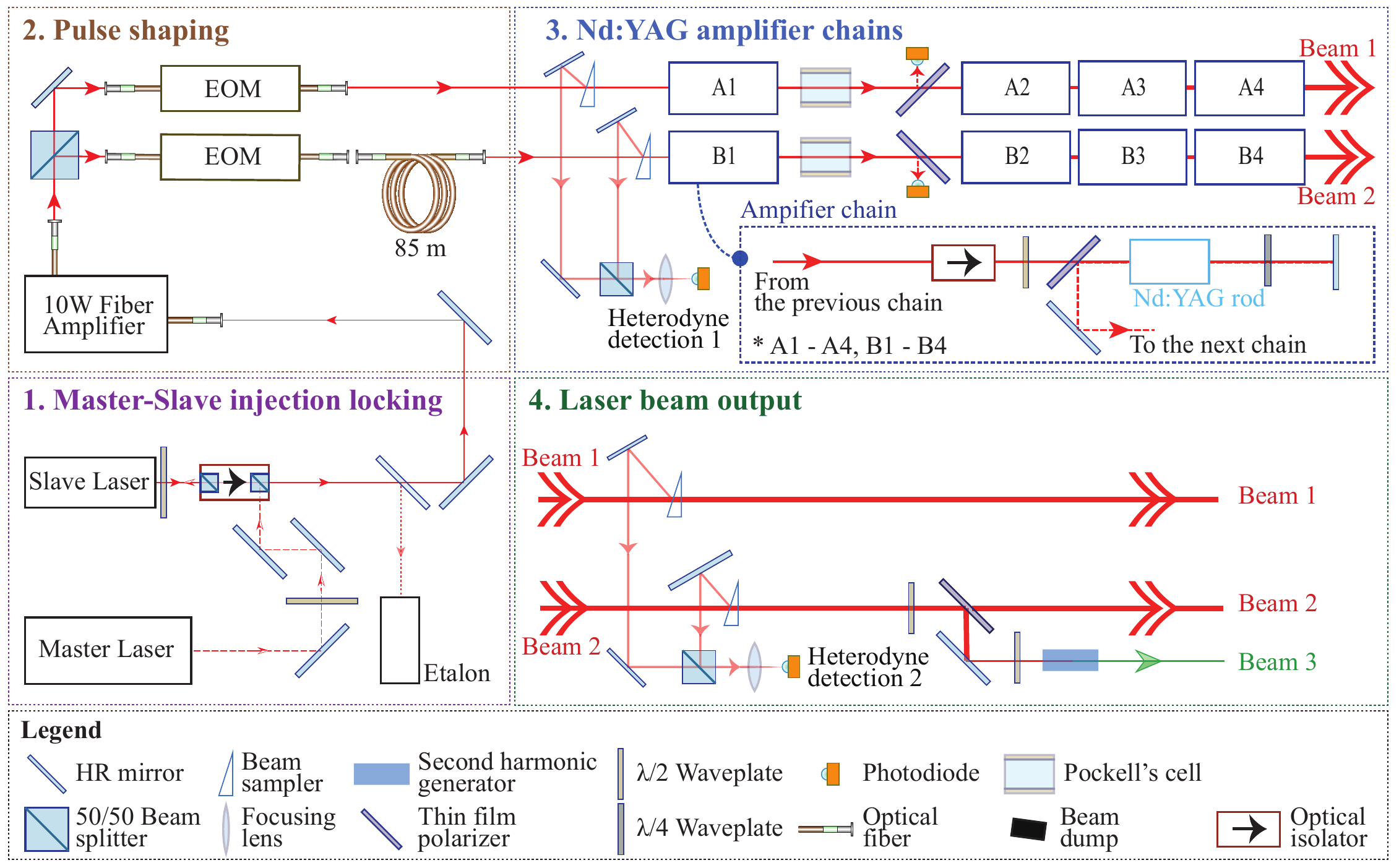}
\caption{\label{fig:lasersys} Conceptual schematic of the dual color, frequency, pulse duration and shape agile laser system presented in this manuscript. The four main sub-systems of the laser are noted in Arabic numbers while the thickness of the line indicating the laser beam path is proportional to the laser energy/power of the beam at that particular point.}
\end{figure}

The overall optical layout of the laser system which features dual color, frequency and pulse duration agile capabilities is presented in Fig.~\ref{fig:lasersys}. A short overall description of the laser's operation will be provided first, before breaking down into the respective laser sub-systems, as these are noted in Fig.~\ref{fig:lasersys}. 

The laser chain starts with a custom designed and constructed, continuous wave (CW) $1064$ nm beam emitted from an Nd:YVO$4$ microchip laser, whose frequency can be rapidly chirped by an intracavity LiTaO$_3$ electro-optic crystal. This laser, which is referred to as the \textit{master}, is injection-locked onto a diode laser, termed the \textit{slave}, which produces a stable intensity CW beam while retaining the frequency information from the master laser. After its power is amplified to $\sim1$ W by a commercial fiber amplifier, this CW beam is split in two beams. Each beam is independently fed into a respective intensity electro-optic pulse shaper where the beam can be chopped up into pulses ranging from a temporal duration of $\sim1$ ns to $\sim1$ $\upmu$s with any arbitrary temporal shape, driven by an arbitrary waveform generator (AWG). These output optical pulses are each amplified by separate diode-pumped Nd:YAG amplifier chains, eventually producing two main $1064$ nm pulsed beams, each of $\sim450$ mJ energy per pulse. Part of one of the main beams is picked up to generate the third beam, which can be frequency doubled to generate a $532$ nm beam for the dual color configuration, maintaining the exact temporal characteristics as the first harmonic. 

The described overall functions of the system are realized mainly by four sub-systems, as noted on Fig.~\ref{fig:lasersys}: 1) Master-slave injection locking, 2) Pulse shaping, 3) Nd:YAG amplifier chains and 4) laser beam output. Each sub-system of the laser system performs specific roles in the laser parameter configuration. The details of the laser sub-systems are discussed in detail in the following sections.

\subsection{Master-slave injection locking}
\subsubsection{A custom-built Nd:YVO$_4$ microchip master laser}

A neodymium based dioded-pumped solid state (DPSS) laser serves as a master laser in the developed system (Section 1 in Fig.~\ref{fig:lasersys}). The main components of the master laser cavity are an Nd:YVO$_4$ crystal, 
acting as the lasing medium\footnote{$2.95$x$3.05$x$0.5$ mm $3$\% Nd doped, AR coated for $808$ nm on the incident side and HR coated for $1064$ nm on the cavity side; CASIX} and a plano-concave partially reflecting (PR) output coupler\footnote{$95\%$ PR@$1064$ nm on the cavity side; LAYERTEC} (OC). The Nd:YVO$_4$ crystal is optically pumped by an $808$ nm pumping diode\footnote{L$808$P$1000$MM; Thorlabs}. Most importantly, an intracavity LiTaO$_3$ electro-optic modulator\footnote{$1$x$3$x$5$ mm, AR coated for $1064$ nm on both laser propagation sides, gold plated on the two orthogonal sides; CASIX} (EOM), comprises almost the whole length of the laser cavity: by applying voltage to the crystal, its refractive index is changed and thus the cavity length, allowing for a fast change of the output laser frequency. 

\begin{figure}[h!]
\centering
\includegraphics[width=0.7\textwidth]{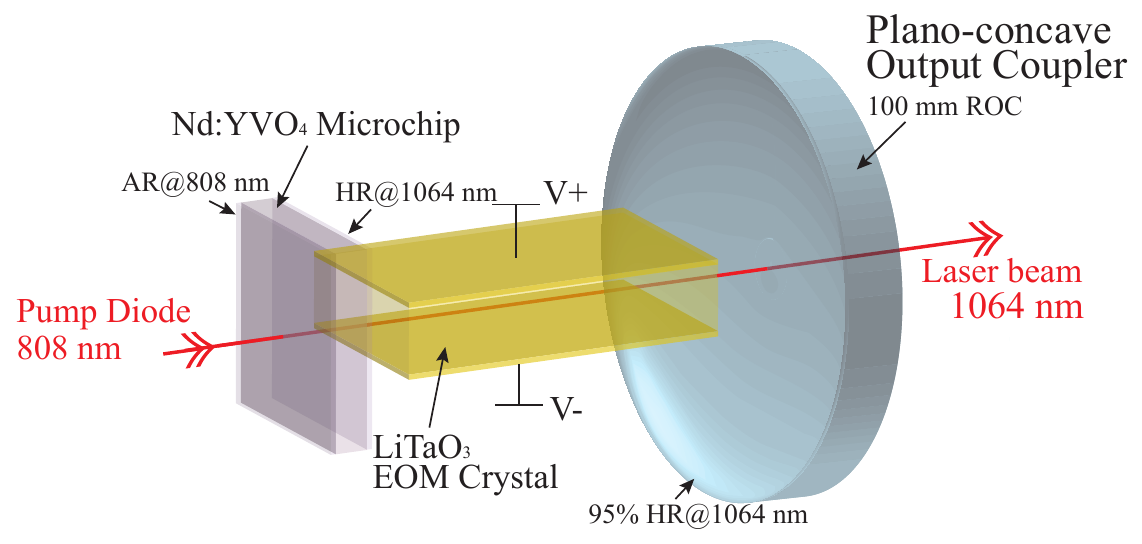}
\caption{\label{fig:cavity}Schematic of the microchip laser cavity.}
\end{figure}

Figure~\ref{fig:cavity} presents a schematic of the microchip laser cavity of the master laser. A shorter laser cavity for the Nd:YVO$_4$ microchip DPSS laser is chosen, since it allows for a larger free spectral range (FSR) of the laser, thus ensuring single mode operation of the laser over the chirped duration of the final laser pulse. Apart from it being readily commercially available, the Nd:YVO$_4$ crystal is preferred as the active lasing medium over an Nd:YAG crystal since it has a higher stimulated emission cross-section and experiences strong, broadband absorption at $808$ nm\cite{Koechner1988PropertiesMaterials}. Additionally, due to its large absorption bandwidth, it is more tolerant to temperature variations of the pumping diode, enabling overall a more stable laser operation. It is noteworthy that a plano-concave (PC) OC is chosen for the cavity design: compared to a more common plano-parallel configuration which fully relies on thermal lensing for lasing, the PC OC allows for the lasing condition to fall well within the stable region of the stability criterion for the laser resonator. The cavity stability criterion is given as $0 \le g_1 g_2 \le 1$, where $g = 1-L/R$ with $L$ representing the cavity length and $R$ the radii of curvature (ROC) of the cavity surfaces, the subscripts representing each cavity forming surface~\cite{Koechner1988PropertiesMaterials}. With the flat surface on the Nd:YVO$_4$ side ($R_1 = \infty$), the PC OC equaling 100 mm ($R_2$ = 100 mm) and the cavity length equaling the length of the intracavity crystal ($L\approx5$ mm): $g_1 = 1$ and $g_2 < 1$, thus satisfying the stability criterion, ensuring stable laser operation.

\begin{figure}[h!]
\centering
\includegraphics[width=0.7\textwidth]{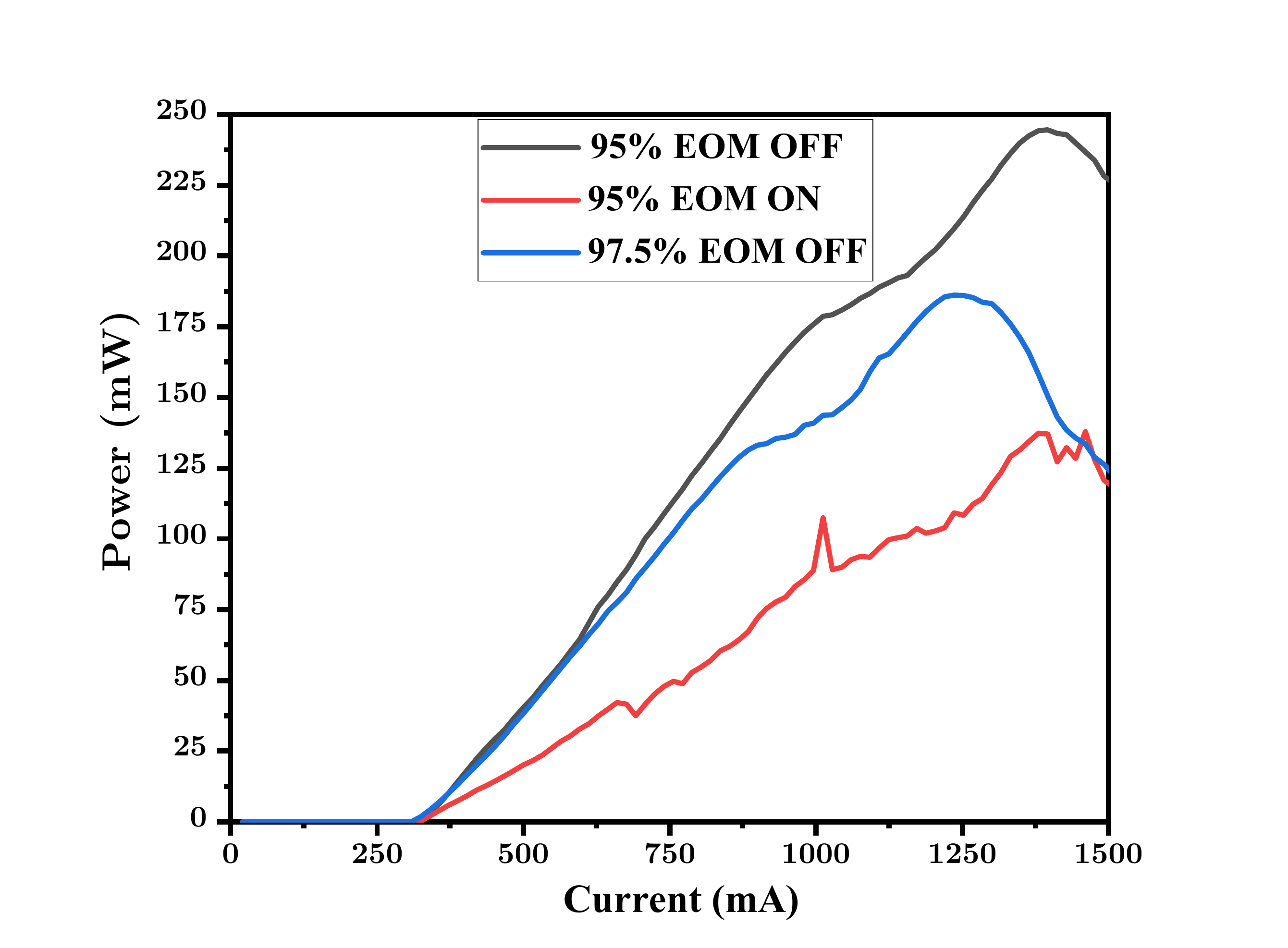}
\caption{\label{fig:IP}Characteristic master laser output power versus pump diode current IP curves for two different output coupler partial reflections: $95$\% (black) and $97.5$\% (blue). EOM ON/OFF indicates whether the EOM is in place or not. The IP curve for when the EOM was placed inside the cavity is shown in red. All curves are obtained for the pump diode at 14 \textcelsius~ which is a nominal temperature.}
\end{figure}

To optimize the performance of the DPSS laser, two OCs having different partial reflectivities (PR) of 95\% and 97.5\%, respectively, were tested. An experimental test like this is preferred over an accurate simulation method for such DPSS lasers as it is much faster and more economical, requiring a direct characterization of the laser performance with actual parts in place. Figure~\ref{fig:IP} shows characteristic I-P curves of the output optical power (P) of the master laser versus the $808$ nm pump diode current (I). Initially, without the EOM in place (i.e. without an extra loss mechanism), the I-P efficiencies for $95$\% and $97.5$\% PR OCs are obtained, leading to the selection of the 95\% PR OC. Additionally, the final master laser efficiency with the $95$\% PR OC and the EOM in place was also measured. All measurements were performed when the laser cavity was aligned to produce a TEM$_{00}$ beam. 

One of the important roles of the master laser in this laser system is to enable frequency chirping, i.e. for relative frequency excursions between the two eventually produced laser beams of the system. The frequency chirping is achieved by the intracavity LiTaO$_3$ EOM crystal. An electrical voltage via a waveform generator\footnote{3314A; Agilent} is amplified by a voltage amplifier\footnote{A-304; A.A. Lab system}, and this amplified voltage is applied onto the gold plated top and bottom side of the crystal to induce a change in the refractive index $n_1$ of the crystal with respect to time, and hence in its optical pathlength $n_1l_1$, where $l_1$ is its physical length. The applied voltage waveform can be of any waveform shape achieved by the waveform generator; usually a linear ramp is preferred, since this is translated to a linear chirp across the final laser pulse duration. The applied voltage on the intracavity crystal effectively equates to a change in the cavity optical path length $l$ and thus a change to the output laser frequency $f$, eventually leading to the laser frequency chirping. The achieved frequency change $\delta f$ in the master laser output with respect to the applied voltage $\delta V$ is given by\cite{Repasky2002}:
\begin{equation}
    \frac{\delta f}{\delta V} = \frac{f n_1^2 r_{33}}{2 d} \frac{n_1 l_1}{l_\mathrm{cav}},
    \label{eqn:chirprate}
\end{equation}
where $f$ is the optical frequency in the cavity, $n_1$ and $l_1$ are the refractive index and length of the LiTaO$_3$ crystal, and $l_\mathrm{cav}$ is the total optical path length given as $l_\mathrm{cav} = n_1 l_1 + n_2 l_2 + n_3 l_3$ where the subscripts 2 and 3 are for the Nd:YVO$_4$ and the air gap, respectively. Thus, as the total optical pathlength of the intracavity LiTaO$_3$ EOM crystal within the cavity increases, the frequency chirping rate (measured in MHz/V) increases. 

\color{black}
\begin{figure}[h!]
\centering\includegraphics[width=0.6\textwidth]{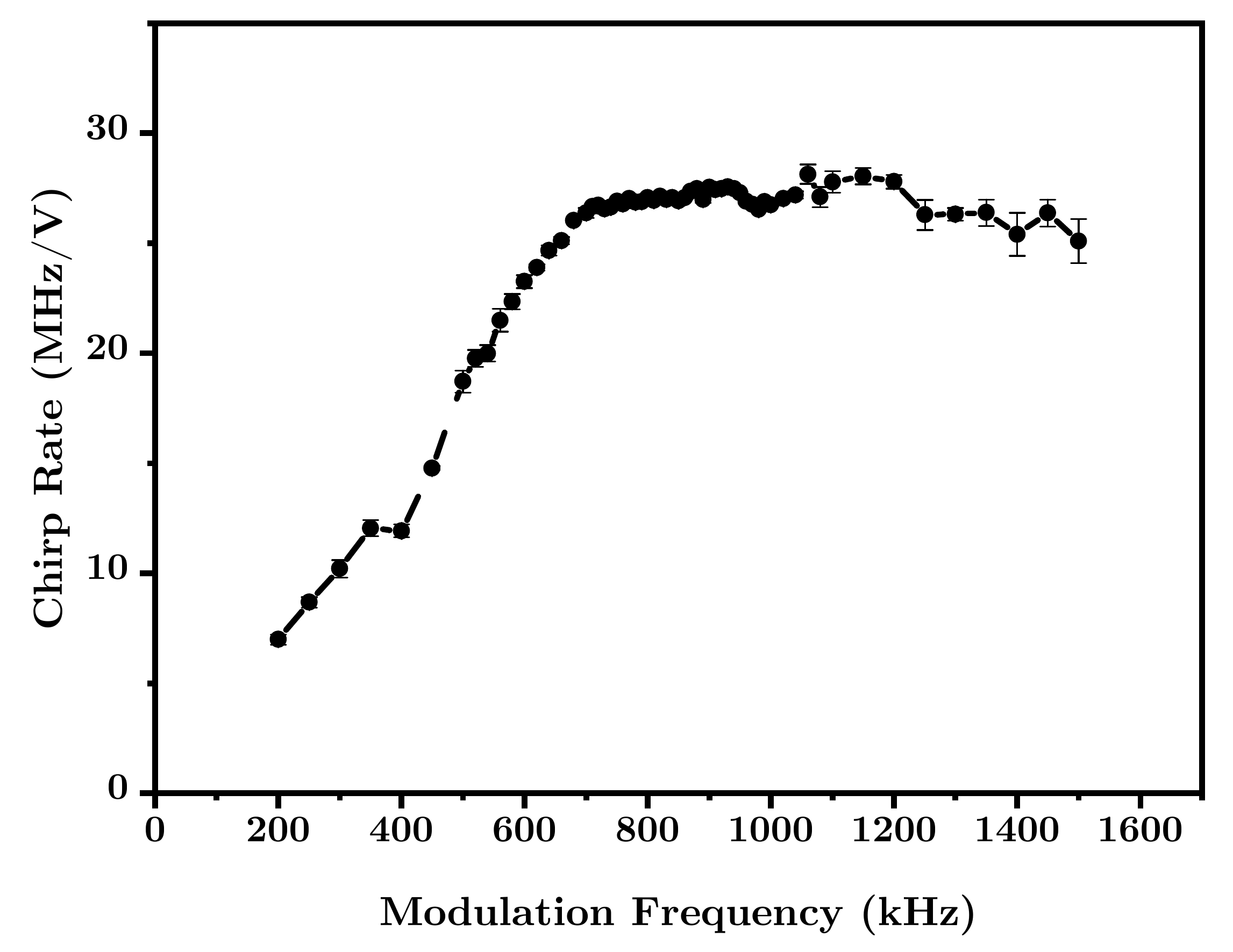}
\caption{\label{fig:chirpmod} Frequency chirp rate (in MHz/V) as a function of the modulation frequency of the applied voltage. The measurement was performed by applying a continuous, sinusoidal voltage of varying voltage amplitude and frequency on the crystal, and measuring the laser output through a Fabry-Perot interferometer.}

\end{figure}

The frequency chirp rate of the present system as a function of the modulation frequency of the electric voltage applied to the intracavity LiTaO$_3$ EOM crystal was measured and is shown in Fig.~\ref{fig:chirpmod}. It should be noted that the modulation frequency referred to here is distinct from the laser (optical) frequency $f$ that is used when estimating the chirp rate. The modulation frequency is the frequency of the sinusoidal voltage signal sent across the EOM crystal, or the inverse of the period for a single chirp. The chirp rate, as defined in Eq.~(\ref{eqn:chirprate}), is the change in the laser frequency over the applied peak-to-peak voltage on the EOM crystal. A high chirp rate is desirable in order to achieve the widest frequency difference between the two eventual laser beams, typically on the order of a few gigahertz; it is favorable if this is achieved using the lowest voltage possible. To quantify the achieved chirp rate of this laser system, a Fabry-Perot interferometer is employed\footnote{SA200-8B, $1.5$ GHz bandwidth; Thorlabs}. The $\delta f$ measured for each individual voltage is proportional to the temporal width of the Fabry-Perot signal over its scanning window, measured as the full width half maximum (FWHM). The FWHM was measured for various peak to peak applied voltages and a linear fit performed, resulting in the chirp rate per modulation frequency. This process was repeated over various voltage modulation frequencies to find the optimal operating modulation frequency. 

At low modulation frequencies, the chirp rate was found to steadily increase with higher modulation frequency, , increasing the rate at which the applied voltage changes the laser frequency. At approximately $700$ kHz modulation frequency the chirp rate starts to reach a plateau, reducing the effect of the modulation frequency on the chirp rate. Higher modulation frequency values lead to larger uncertainties in the chirp rate.
At the optimal operating conditions, determined through the process described above, a chirp rate of approximately $27$~MHz/V is obtained. For a $\pm 50$~V voltage sweep, this corresponds to an achievable frequency scanning range of $\pm 1.3$~GHz. This is approximately double the previously reported frequency chirping rate of $\sim15$~MHz/V for a similar system\cite{Gerakis2013}. Given the fact that we can supply up to $\pm 400$~V voltage sweep within the same pulse duration, we estimate that a full chirp range exceeding $10$ GHz can be achieved with this system. For this to be verified, a Fabry-Perot interferometer with a larger free-spectral-range (e.g. at $10$ GHz) is needed.

We note that such rapid frequency chirping in a Nd based solid-state laser can induce relaxation oscillations\cite{Siegman1986Lasers}, which can lead to undesirable, strong laser output intensity oscillations. The relaxation oscillations are less strong with an Nd:YVO$_4$ crystal, since it has a shorter upper-state lifetime versus e.g. an Nd:YAG crystal\cite{Koechner1988PropertiesMaterials}. Nevertheless, the relaxation oscillations would cause severe intensity fluctuations at the final output of the laser system. To resolve this undesirable effect, a master-slave injection locking scheme was implemented and is described in the next section.

\subsubsection{Slave laser}\label{sec:slave}
As mentioned in the previous section, with a master-slave injection locking scheme, undesirable laser intensity fluctuations due to the relaxation oscillations are avoided. This is because only the frequency information of the master laser is transmitted to the slave laser, while the intensity fluctuations are disregarded if the master laser's input power exceeds the slave laser's saturation power. When the frequencies of both the master and slave lasers are close enough, the chirped master laser forces the slave laser to emit at the exact same frequency through a strong, saturated coupling while maintaining a stable output intensity. 

A $1064$ nm diode laser\footnote{M9-A64-0200; Thorlabs} is used as a slave laser with benchtop controllers\footnote{LDC205C and TED200C; Thorlabs} controlling its operating current and temperature. Temperature tuning of the laser diode is a crucial step in the process, since it enables the laser diode to reach an emitting wavelength close to that of the master laser: this is a key to a successful injection locking scheme. To prevent any back-reflection of the slave laser output into itself, an optical isolator\footnote{110-21022-0001-C1; Electro-Optics Technology} is placed in front of the slave laser. To maximize the coupling between the master and slave lasers, the output of the master laser is fed into the slave laser through a rejecting port of the optical isolator (see Fig.~\ref{fig:lasersys}). This improves the total transmission of the master laser into the slave laser. The status of the injection locking is monitored by a Fabry-Perot interferometer, which monitors the characteristic shape of the power spectrum of the injection-locked laser\cite{Troger1999Frequency-sweepLocking}.

\subsection{Temporal Pulse Shape Manipulation}
The frequency chirped CW laser output from the master-slave section is sent to the pulse shaping section of the laser system. Prior to being shaped into laser pulses, the CW beam coming from the master-slave section is initially further amplified by a commercial, Yb-based fiber amplifier\footnote{YAR-10K-1064-LP-SF; IPG Photonics} that is capable of emitting up to $10$ W power at the same frequency. The amplified output is split in two through a $50/50$ beamsplitter cube, with each resulting beam being fed into a respective commercial Mach-Zender type intensity EOM\footnote{AM1064b; Jenoptik} (iEOM). Each iEOM is used to shape a pre-amplified seed pump pulse. It should be noted that the adaptation of this type of iEOM enables the pulse shaping with commonly available arbitrary waveform generators since the required voltage for shaping is relatively low (range of -$5$V to $5$V). This advantage allows for avoiding the use of high voltage switching equipment such as Pockel's cells, which typically operate in the kilovolt range and induce a strong electromagnetic interference noise.  Pulse shaping begins by shaping the first pulse (Beam 1) with an iEOM, where initially the unchirped, constant frequency part of the laser beam is sent through the iEOM. At a later time, an electrical voltage is applied to the intracavity EOM in the master laser when shaping the second pulse (Beam 2) with the second iEOM: this enables inducing a relative frequency difference between the two pulses. This relative frequency difference is solely determined by the voltage waveform applied to the intracavity EOM. Additionally, since both pulses originate from the same master laser, as long as the master-slave laser system is stable within the hundreds of ns of the pulse shaping operation, the overall operation of the laser system will also be stable. Finally, to temporally align the two produced pulses, the first pulse goes  through an $85$ m optical fiber delay line. 

An arbitrary waveform generator\footnote{AFG3102C; Tektronix} is used to drive the iEOMs to shape and manipulate the pulses from the CW beam. This provides the capability for additional delay control. In addition, and importantly, the arbitrary temporal pulse shaping allows for the generation of a flat top output pulse by accommodating the time dependent gain in the Nd:YAG amplification chains (discussed in Section~\ref{sec:amplifiers}).

\begin{figure}[h!]
\centering\includegraphics[width=\textwidth]{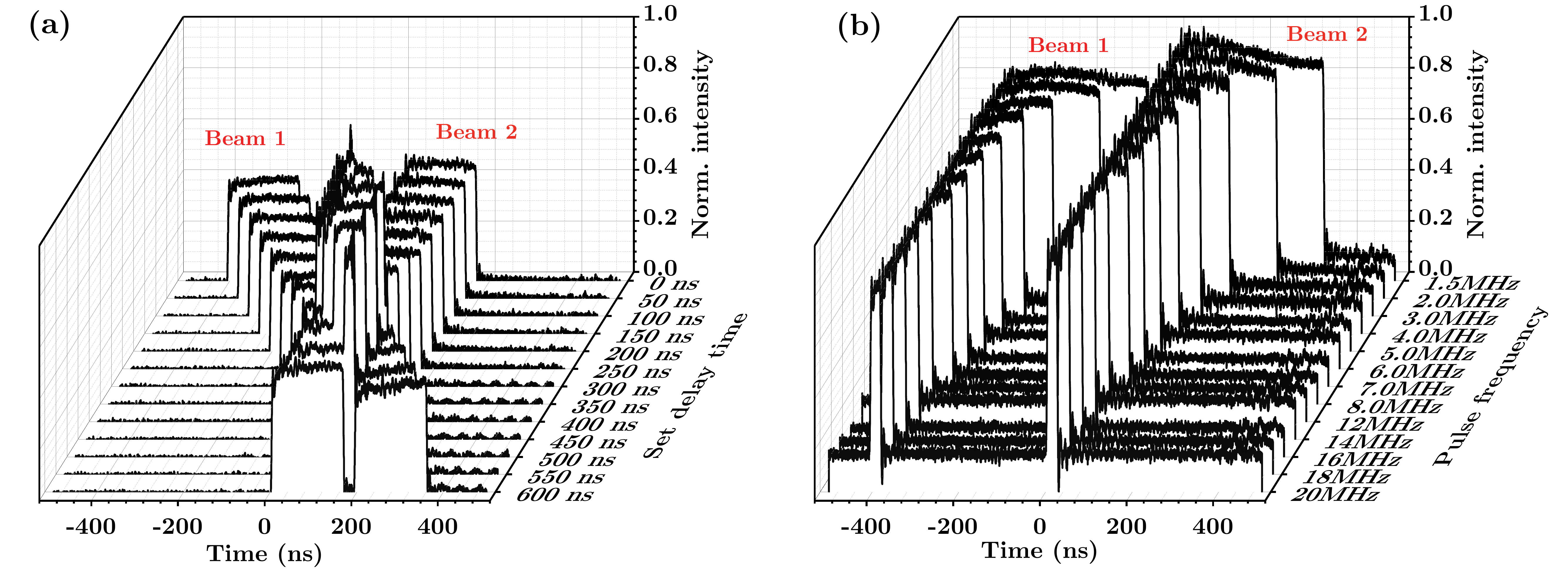}
\caption{\label{fig:pul_mani} Temporal property manipulation of the two pulsed beams: (a) Manipulation of relative time delay between pulses and (b) manipulation of temporal pulse duration.}
\end{figure}

Temporal pulse delay manipulation and pulse duration manipulation examples are shown in Fig.~\ref{fig:pul_mani}. In Fig.~\ref{fig:pul_mani}~(a), the temporal position of Beam 2 is fixed in time, while the relative timing of Beam 1 is changed. In Fig.~\ref{fig:pul_mani}(b), the temporal duration of both pulses is manipulated by controlling the period of the voltage signal applied to the iEOMs. It is noteworthy that each pulse can be individually controlled, both in temporal shape/duration as well as in their relative delay. The obtained curves are taken by a fast photodiode located before the Nd:YAG amplifier chains (heterodyne detection $1$ in  Fig.~\ref{fig:lasersys}). Examples of arbitrary temporal pulse shape manipulation are shown in Section~\ref{sec:laser_output}, where the temporal shapes before and after the second harmonic generation (SHG) are compared (Fig.~\ref{fig:shg_shape}).

The freedom in the pulse delay control from the iEOMs enables the accurate control of the produced beams simultaneously in the test section. This is especially beneficial for applications involving multiple-wave mixing techniques (such as e.g. single shot CRBS) where all beams are required to propagate the scattering volume at the same time. Alternatively, two separate pulsed beams can also serve as individual laser sources for applications such as dual-pulse laser ignition~\cite{Mahamud2018Dual-pulseModel}. The controllable pulse duration is also highly advantageous in applications such as the ones involving strong background optical radiation. In these cases, the short probing pulse can be used to minimize the background collection for better signal-to-noise ratios. On the other hand, a longer duration pulse may be desirable for applications where the high peak laser intensity can be problematic. For instance, a longer duration pulse for the same laser pulse energy can avoid undesirable laser perturbation such as laser heating or photo-ionization, which could be problematic in e.g. Thomson scattering measurements on low density plasmas\cite{Carbone2015ThomsonChallenges}.

\subsection{Nd:YAG amplifier chains}
\label{sec:amplifiers}
Each of the two generated pulsed beams undergoes pulsed amplification by propagating through two arms, each comprising of a total of four diode pumped Nd:YAG amplifier stages in tandem. In each of these arms, the laser pulse at an initial energy per pulse of $\sim$100~nJ is amplified to a final energy per pulse of $\sim$450 mJ after all amplification stages. We note that the diode pumped amplifiers used here\footnote{In tandem in each stage: one RBAT20, two RBAT34, and one REA7006 from Northrop Grumman Cutting Edge Optronics} have several advantages compared to flash lamp pumped amplifiers since they are more compact in size, silent, and have good electrical-to-optical efficiency. Each amplifier stage in the chain is set in a double-pass single-stage amplifier configuration, with each stage consisting of a Faraday optical isolator, a half-wave plate, a thin-film polarizer, a diode-pumped Nd:YAG rod, a quarter-wave plate, and a high reflective mirror for $0\degree$ angle of incidence (refer to section 3 of Fig.~\ref{fig:lasersys}). The Nd:YAG amplifiers in each stage increase the energy from $\sim100$~nJ to $\sim12$~$\upmu$J in the first stage, to $\sim800$~$\upmu$J in the second, $\sim20$~mJ in the third, and finally to $\sim450$~mJ in the fourth, final stage. Additionally, we note that along the amplifier chain, the laser beam diameter is properly adjusted with the use of beam expanders to a beam diameter corresponding to $\sim80$\% of the Nd:YAG rod diameter as recommended by the manufacturer. By selectively operating the respective amplification stages, the final output laser pulse energy is easily controllable while, with the addition of more amplifier stages, the energy per pulse achievable with this system is highly scalable.

\subsection{Laser beam output section}
\label{sec:laser_output}
\subsubsection{First harmonic generation}
Throughout the previously described sections of the laser system, the initial first harmonic CW beam eventually provides two fundamental first harmonic $1064$ nm amplified beams (Beam 1 and Beam 2 in Fig.~\ref{fig:lasersys}) having a Gaussian spatial beam shape with a beam diameter of $\sim5.6$ mm with variable pulse energy and duration, which can be utilized for various applications. For instance, the typical characteristics of the beams used for CRBS application (discussed in Sec.~\ref{sec:apps}) are as follows: a pulse energy of $\sim450$ mJ, a pulse duration of $\sim 150$ ns with a flat-top temporal profile.

\begin{figure}[h!]
\centering\includegraphics[width=1\textwidth]{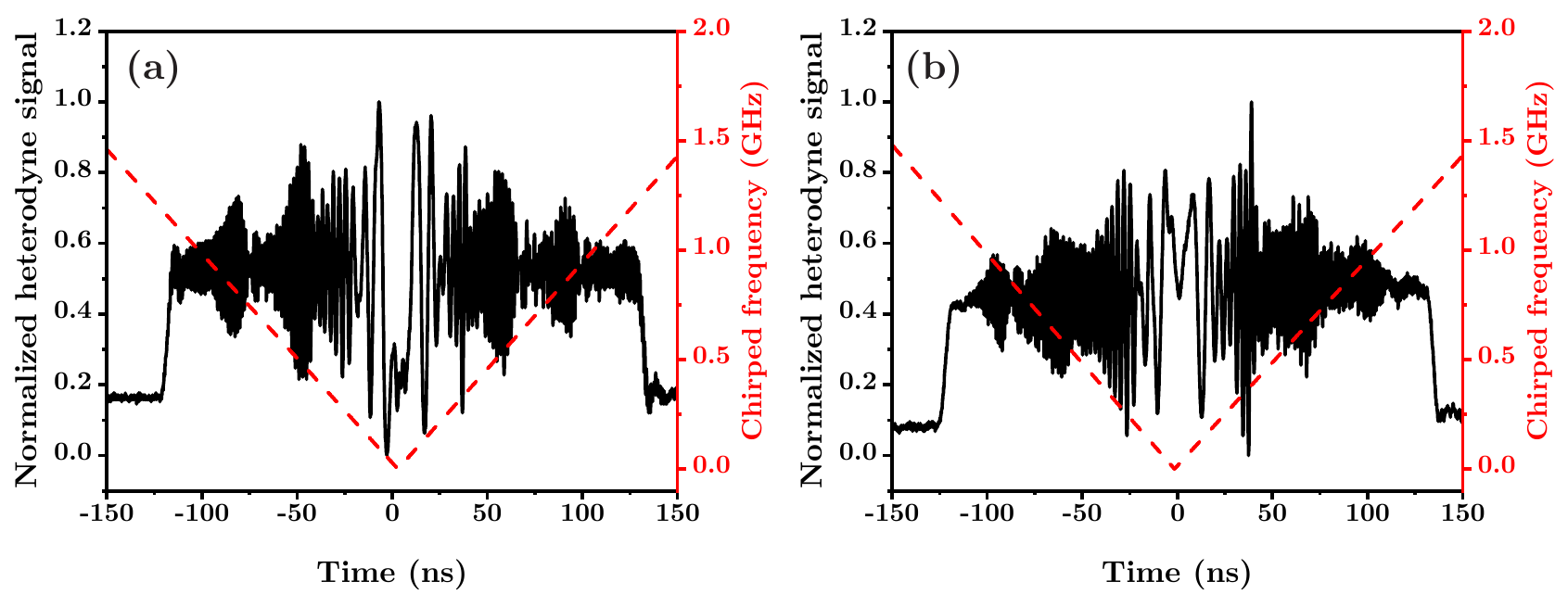}
\caption{\label{fig:hetero} Heterodyne signal (solid black line) and analyzed chirped frequency (red dashed line) are shown for (a) before and (b) after amplification. This result demonstrates that the pulse amplification is not altering the chirp as set by the master laser.}
\end{figure}

Each beam is sampled and heterodyned on a fast InGaAs photodiode\footnote{G6854-01; Hamamatsu Photonics} at the heterodyne detection location (see Section 4 of Fig.~\ref{fig:lasersys}). The heterodyned signal of the two beams provides information on the final, relative chirped frequency between the two beams. Sample heterodyne signals before and after the amplification stages are shown in Fig.~\ref{fig:hetero}. The obtained beat patterns contain the chirped frequency information and are analyzed as described in Ref.~\cite{Fee1992OpticalSpectroscopy}. In Fig.~\ref{fig:hetero}, the extracted chirped frequency based on this method, for the heterodyned signal before and after amplification, is shown. 
In the sample heterodyne signals shown in Fig.~\ref{fig:hetero}, the relative frequency difference between the two beams is initially large, then becomes zero at $t=0$, and then increases again: this practically translates to an optical lattice starting from a high velocity, decelerating to zero relative frequency difference (velocity of $0$ ms$^{-1}$) and accelerating again to a higher velocity.

\begin{figure}[h!]
\centering\includegraphics[width=0.6\textwidth]{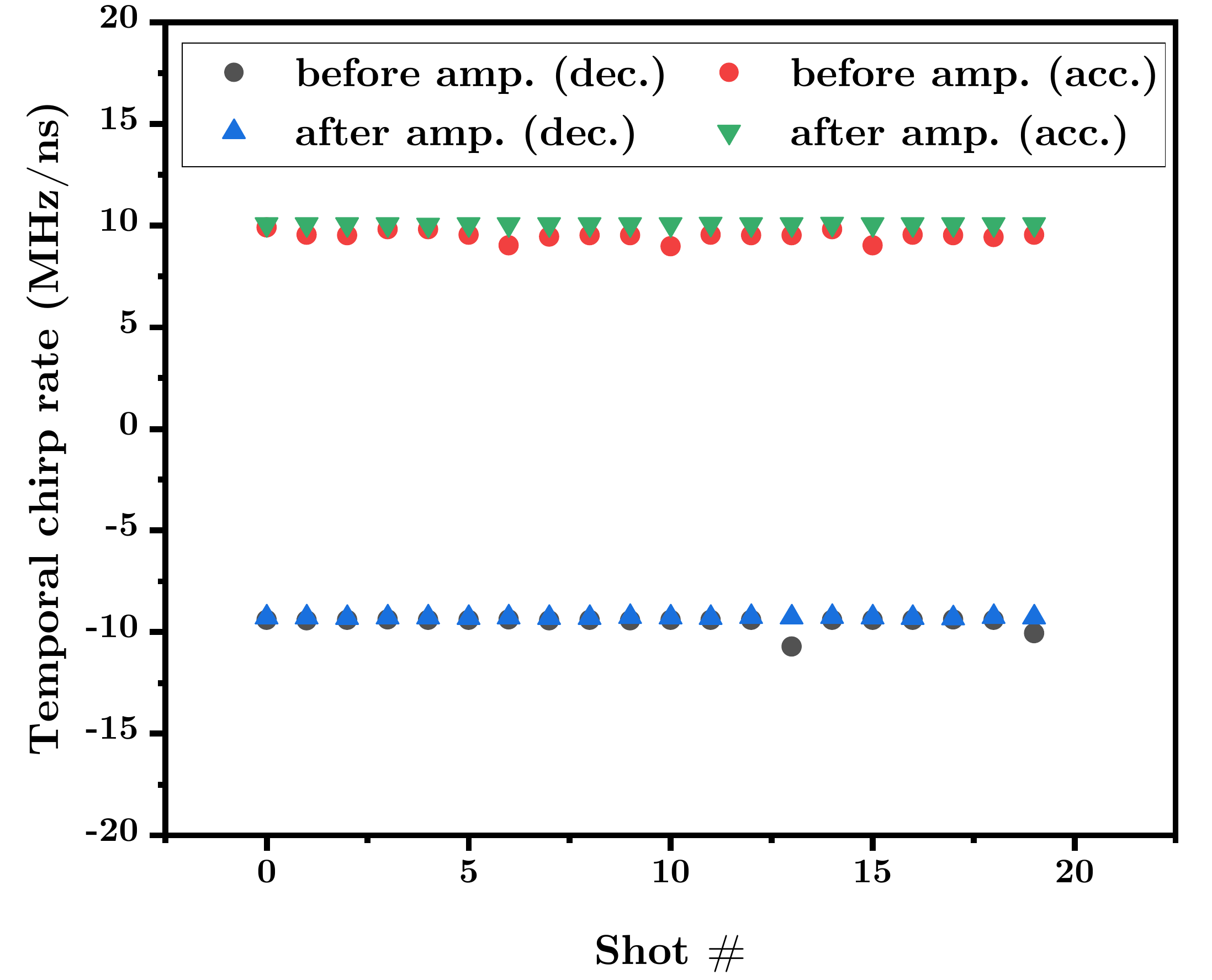}
\caption{\label{fig:chirp} Temporal chirp rate repeatability for 19 total sample shots are shown. Circles are from the pulses before amplification, and triangles are from those after amplification. Temporal chirp rates during both the frequency deceleration and acceleration phases are shown.} 
\end{figure}

The repeatability and stability of the chirp rate before and after amplification is checked and shown in Fig.~\ref{fig:chirp}. The relative standard deviation over the total 19 shots was found to be $<4\%$ before the amplification and $<0.1\%$ after the amplification. The smaller deviation after the amplification is attributed to a better signal-to-noise ratio. The good agreement in both the heterodyne signals and the chirp rates before and after the amplification indicates that the amplification chains do not introduce any disturbance in the achievable chirp rate of the system.

\subsubsection{Second harmonic generation}
\begin{figure}[h!]
\vspace{-5mm}
\centering\includegraphics[width=0.7\textwidth]{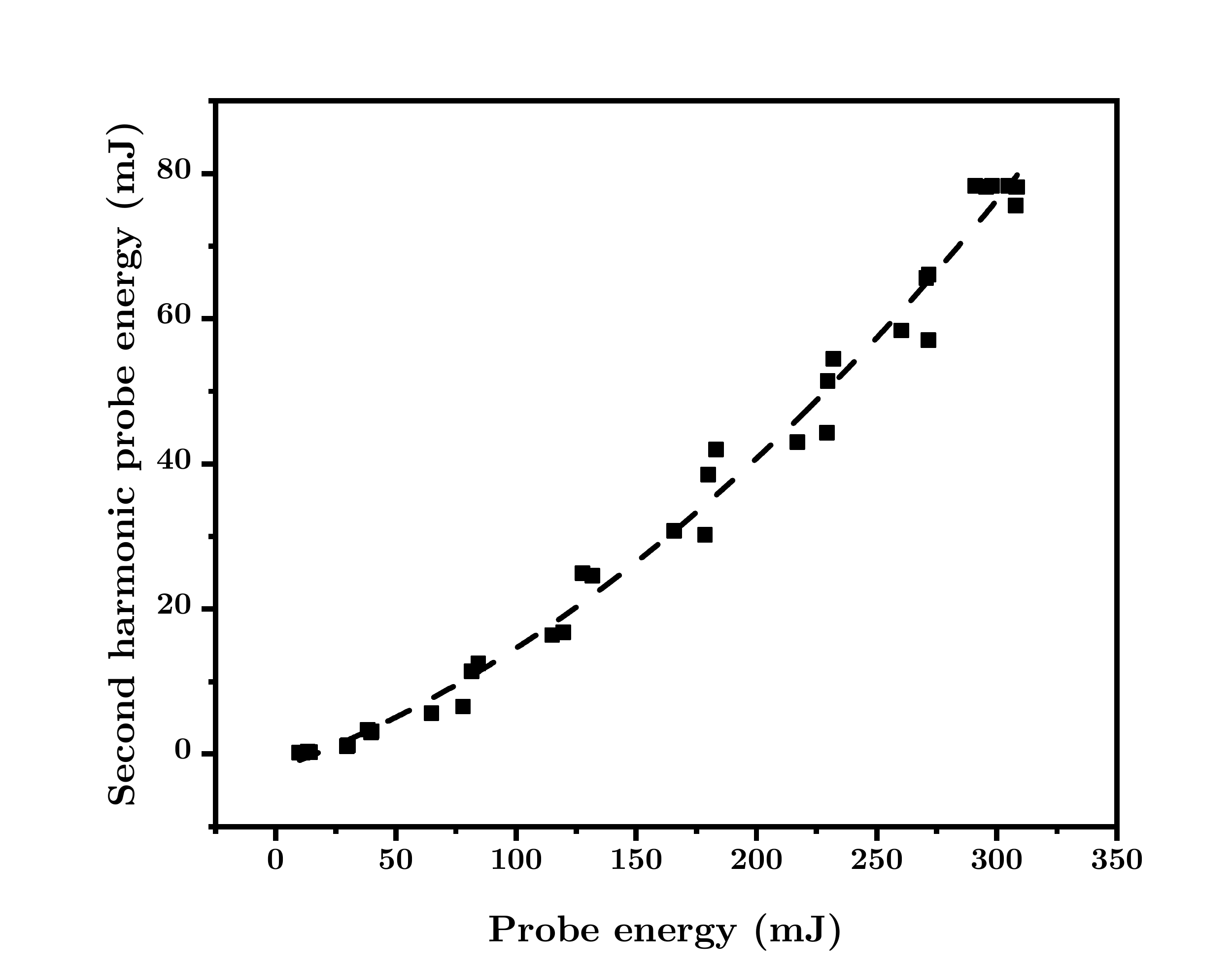}
\vspace{-2mm}
\caption{\label{fig:shg}Second harmonic probe energy as a function of $1064$ nm probe beam. The data points are represented as square points along with the fitted curve for a visual guide.}
\end{figure}

For a dual wavelength configuration, part of Beam $2$ is sampled to produce the second harmonic beam (Beam 3 in Fig.~\ref{fig:lasersys}). The second harmonic beam is generated by a LBO crystal\footnote{$6$x$6$x$50$ mm; AR@$1064$+$532$ nm; Eksma Optics} operating with a commercial crystal oven\footnote{CO1-50-6/6; Eksma Optics} and a temperature controller\footnote{TC2; Eksma Optics}. The second harmonic $532$ nm probe energy per pulse as a function of the $1064$ nm beam input energy per pulse on the crystal is shown in Fig.~\ref{fig:shg}: approximately $27$\% conversion efficiency is obtained. 

It should be noted that, even though there is a natural loss in the achievable energy per pulse, a dual color option can effectively provide a stronger final signal for several applications such as linear laser light scattering measurements. For example, Rayleigh scattered signals scale as $\lambda^{-4}$, where $\lambda$ is the wavelength of the laser used. As a result, the signal for the $532$ nm probe beam will be sixteen times stronger than the $1064$ nm signal at the same energy. Additionally, detectors with a much higher quantum efficiency are available for $532$ nm than those for $1064$ nm. Another practical advantage is that, by adopting a dual color configuration, i.e. $1064$ nm pump beams and a $532$ nm probe beam in any phase matching required applications, the signal beam no longer shares the same optical path with any of the pump beams, which is often the case in a single color co-planar phase matching configuration; an example is single color coherent Rayleigh-Brillouin scattering (CRBS) versus dual color CRBS. All these advantages lead to a significant improvement of the achievable  signal-to-noise ratio. 

\begin{figure}[h!]
\centering\includegraphics[width=1\textwidth]{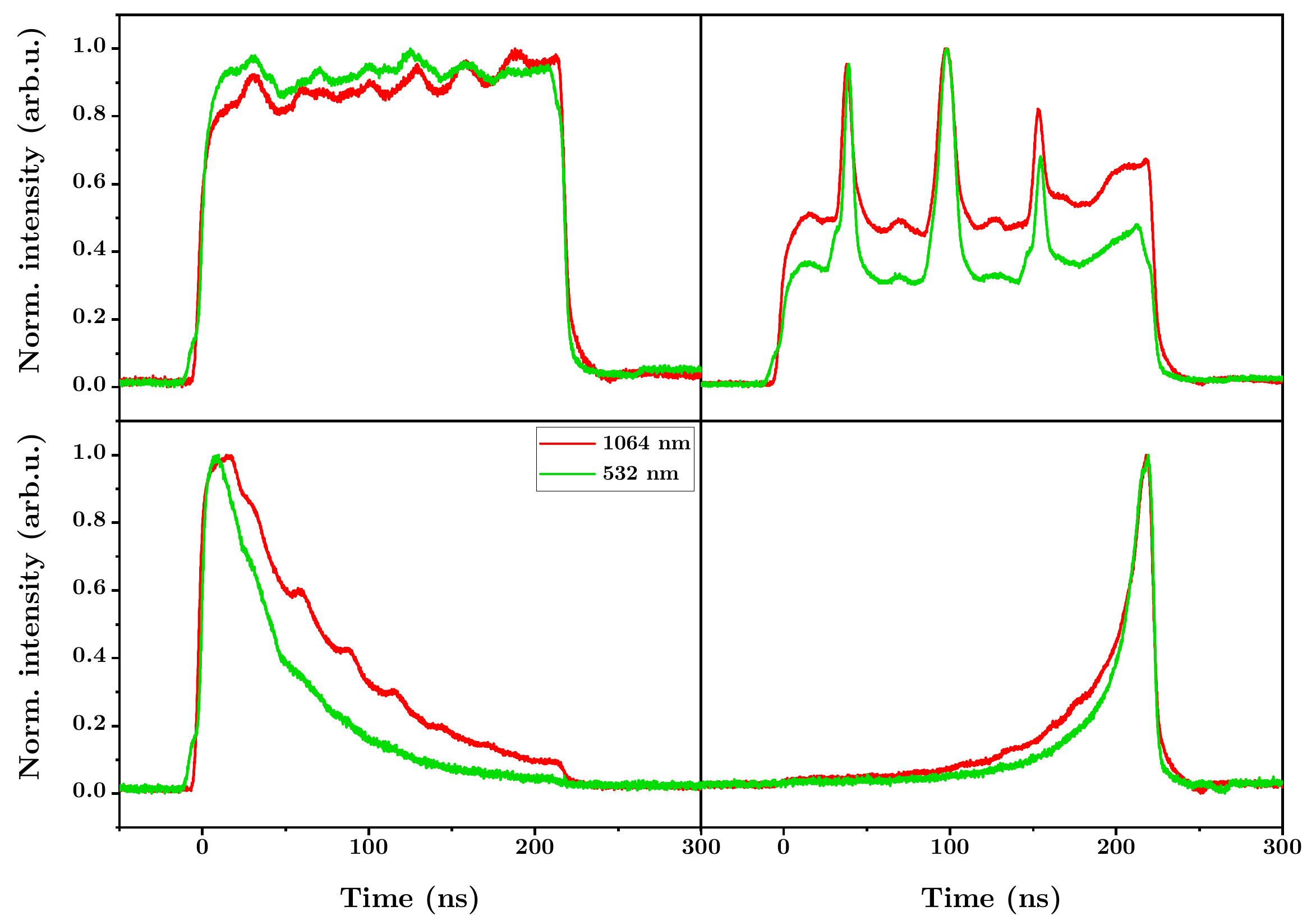}
\caption{\label{fig:shg_shape}Arbitrary temporal shapes of the third beam generated by the laser system presented here. Both a $1064$ nm beam, before the second harmonic generation stage (red line), and a $532$ nm beam, after the second harmonic generation stage (green line), are shown. Note that the slight differences on the temporal shapes is attributed to the different sensitivities of the photodiodes used for the respective wavelengths.}
\end{figure}

\begin{figure}[h!]
\centering\includegraphics[width=\textwidth]{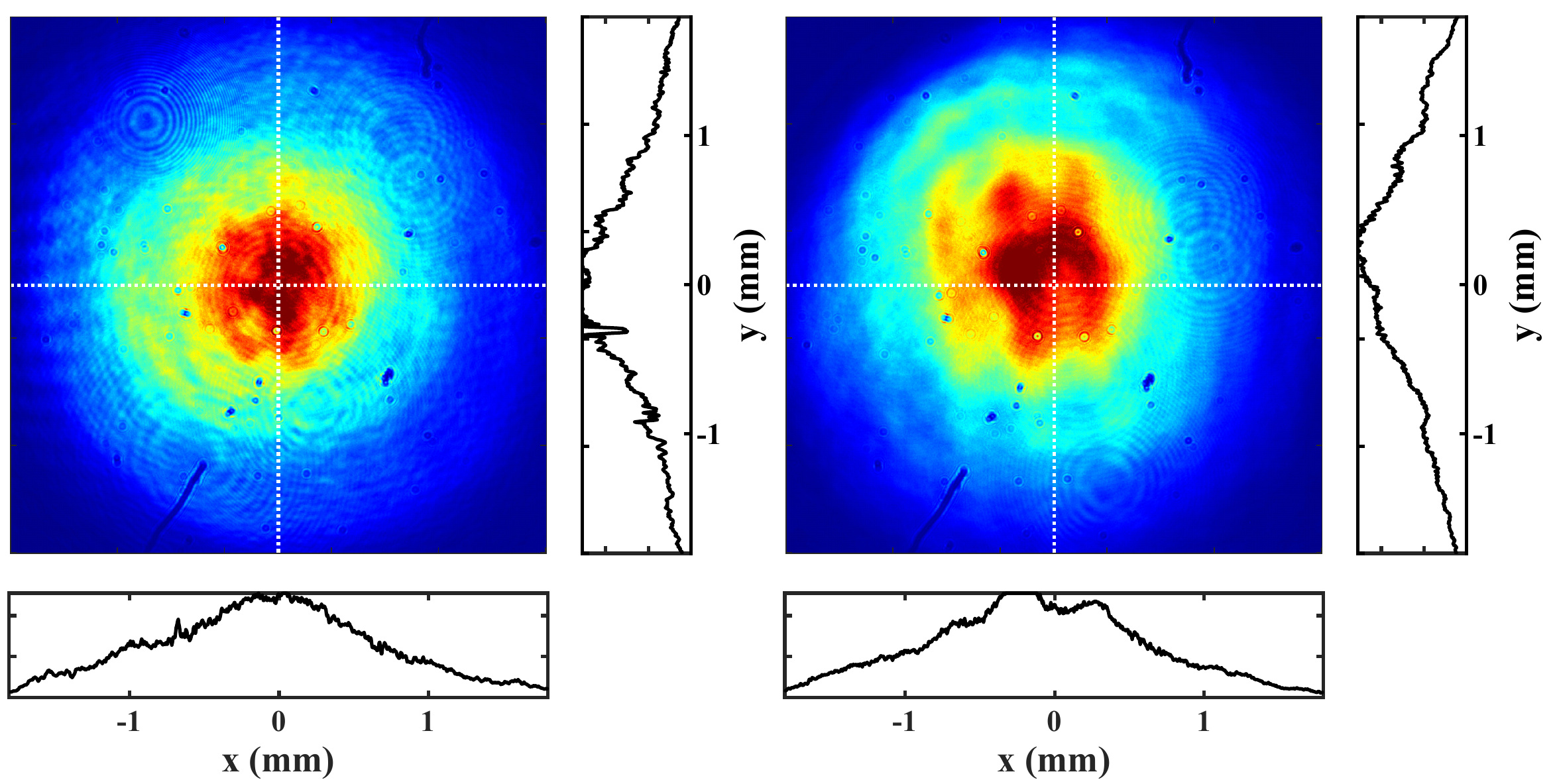}
\caption{\label{fig:shg_profile}Spatial beam profiles of (a) the 1064 nm probe beam and (b) the second harmonic 532 nm. Horizontal and vertical distributions in mm (along the white dashed lines) are given on the bottom and right sides.}
\end{figure}

Importantly, we demonstrate here that the generated Beam $3$ at $532$ nm can maintain the temporal characteristics of the first harmonic. Figure~\ref{fig:shg_shape} shows several sample arbitrary temporal shapes for the second harmonic probe beam compared with the $1064$ nm first harmonic, normalized by the peak intensity. It is notable that the temporal shape is well preserved, thus enabling great flexibility on pulse temporal manipulation for both harmonics. The slight difference on the observed temporal shapes is attributed to the different sensitivity of the photodiodes used for the respective wavelengths. Finally, we note that both spatial beam profiles have good Gaussian shapes as shown in Fig.~\ref{fig:shg_profile}.

\section{Applications of the laser system}\label{sec:apps}
The flexibility and capabilities of the laser system presented here are expected to be useful and potentially enabling for a variety of laser-aided diagnostics and optical manipulation experiments. Particularly, applications with optical lattices~\cite{Greiner2008OpticalLattices, maher2012laser, PhysRevLett.93.243004, Madsen_2021, PhysRevA.100.023401} that are created by the interference of two laser beams can benefit from the rapid frequency chirping capability provided with this all-optical setup, in comparison to the mechanical setups that are usually the current state-of-the-art. Additional applications can also include gas diagnostics such as single shot CRBS\cite{Gerakis2013,Gerakis2018,Gerakis2021}, micropropulsion by molecular manipulation\cite{Shneider2006}, optical microlinear deceleration and acceleration\cite{Barker2001,Barker2002SlowingDeceleration}, and particle spectroscopy\cite{Gerakis2022RamanLattices}. With a controllable pulse duration, the present system coupled with gated detection may benefit applications with accompanying strong background radiation such as spark or arc discharges\cite{Yatom2017DetectionIncandescence,Gerakis2018}. In addition, the system could be of benefit to applications where a longer pulse is desirable but the peak laser intensity is problematic, such as laser heating\cite{Carbone2012RevisionLAPD} and photo-ionization\cite{Carbone2015ThomsonChallenges}.

With regard to CRBS in particular,, the flexibility of the pulse duration and chirping rate can accommodate a wider range of spectral lineshape changes resulting from temperature and pressure variations~\cite{Liang2017} or from the existence of a bulk flow velocity~\cite{Gerakis2021,Bak2021AchievingScattering}. Additionally, the dual color single shot CRBS configuration using a $532$ nm probe beam will enable access to gas information at much lower pressure applications than ever reported in the past~\cite{Graul2014,Pan2002b,Vieitez2010CoherentMixtures,Wang2020}. Furthermore, each beam produced by the system presented here could be utilized for a plethora of different applications such as laser induced plasma formation~\cite{LeDrogoff2004InfluenceProperties}, surface ablation~\cite{Amoruso1997AbsorptionPlasmas}, and dual-pulse laser ignition applications\cite{Shneider2011TailoringPulse,Mahamud2018Dual-pulseModel}, among others.

\begin{figure}[h!]
	\centering\includegraphics[width=\textwidth]{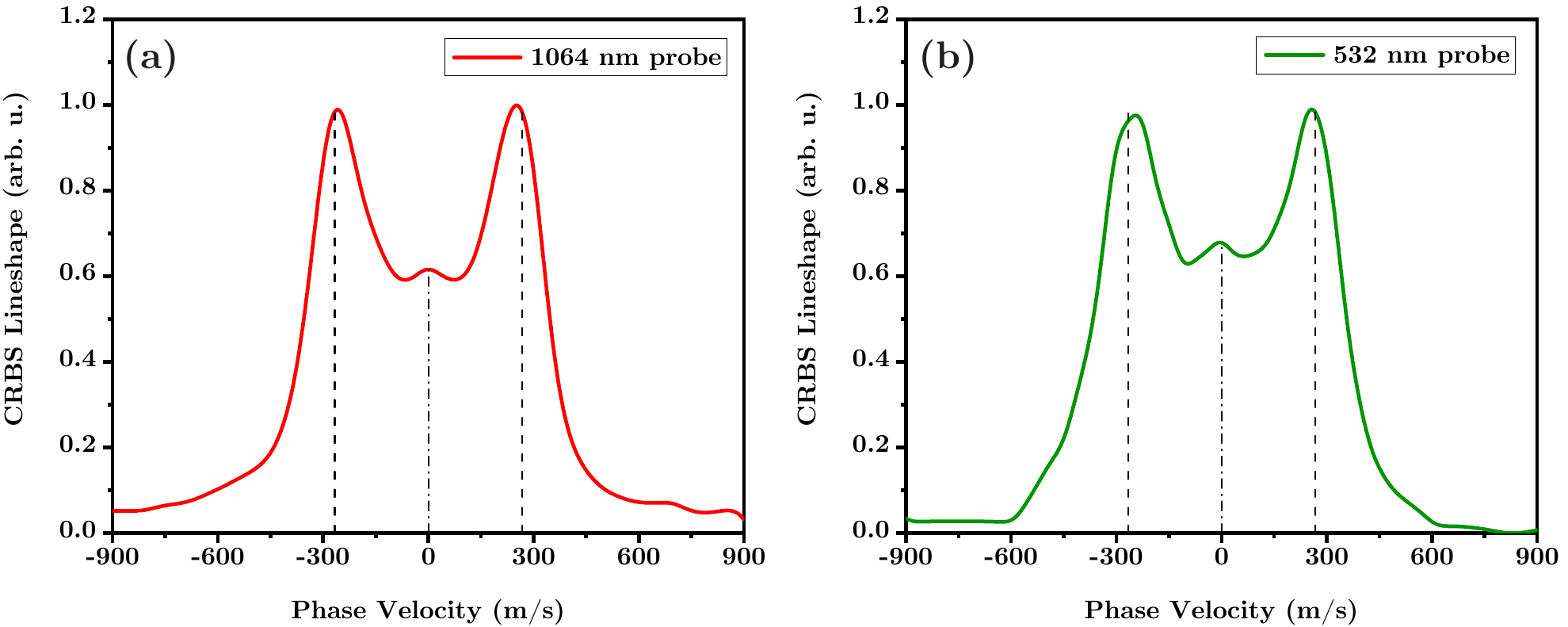}
	\caption{\label{fig:CRBScompa}Sample experimental CRBS lineshapes of CO$_2$ gas at room temperature and atmospheric pressure; (a) with a $1064$ nm probe beam and (b) with a $532$ nm probe beam. The dashed-dot line indicates the Rayleigh peak centered at $0$~m/s, and the black dashed line indicates the speed of sound for CO$_2$ ($267$~m/s at $293$~K and pressure of $1$~atm, estimated by the sound speed equation $v_\mathrm{sound}=\sqrt{\gamma R T}$, where $\gamma$ is the adiabatic constant, $R$ is the specific gas constant, and $T$ is the gas temperature).} 
\end{figure}

As a demonstration of the capabilities of present laser system, we utilize all beams produced by the system to obtain single shot CRBS lineshapes, in single and dual color CRBS geometries. In this application, the two frequency chirped pump beams at $1064$ nm are crossed in air at an angle of $\sim178\degree$ to generate temporally decelerating and accelerating optical lattices within the $\sim150$ ns duration of the pulse: this enables the interaction of the optical lattice with the populations in all velocity domains of the velocity distribution function, within a single laser shot. 
The third probe beam, either at $1064$ nm or at $532$ nm, is incident on the lattice at the Bragg angle ($89\degree$ and $\sim30\degree$ respectively for $1064$ nm and $532$ nm), generating a fourth signal beam containing the single shot CRBS lineshape that provides thermodynamic information of the gas properties, such as temperature, density and flow velocity. Figure~\ref{fig:CRBScompa} shows sample CRBS lineshapes of CO$_2$ gas at room temperature and atmospheric pressure, where (a) is obtained with a $1064$ nm probe beam, and (b) is obtained with a $532$ nm probe beam. Note that the lineshapes are averaged over five shots. The dashed-dot line and the black dashed line indicate the Rayleigh peak centered at $0$~m/s and the speed of sound in CO$_2$ ($267$~m/s at $293$~K and pressure of $1$~atm), respectively. Successful CRBS lineshape acquisition with the $532$ nm probe suggests the accessibility toward much lower pressure (gas density) regimes for gas property diagnostics, using the dual-color single shot CRBS configuration. This can in turn enable advances e.g. in low pressure plasma applications for determining departures from local thermodynamic equilibrium condition in a plasma, or for flow velocimetry in a hypersonic wind tunnel. 

\section{Conclusion and future directions}
In this manuscript, a dual color, frequency and pulse duration agile laser system is presented. The system consists of four subsections where specific laser parameters are configured: master-slave injection locking (frequency manipulation), pulse shaping (pulse temporal duration and shape manipulation), Nd:YAG amplifier chains (pulse energy manipulation), and laser beam output (second harmonic generation). In summary, the system is capable of (i) achieving variable pulse durations ranging from $\sim1$ ns to $\sim1 \upmu$s with (ii) arbitrary temporal profiles at (iii) a dual wavelength output configuration, (iv) exhibiting faster chirping rates to what has been reported with chirp range up to $10$ GHz, and (v) with scalability of pulse energies from some nJ per pulse up to Joules per pulse. These capabilities of the presented laser system are demonstrated by obtaining single shot CRBS spectra in both single color and dual color configurations.
It is envisioned that the laser system could be of benefit to various applications for spectroscopy, gas and plasma diagnostics as well as particle manipulation, as one can tailor the specific laser parameters listed above with ease, as would be required by each application. 

\section{Acknowledgements}
This material is based upon work supported by the National Science Foundation under Grant No. $1903481$.

\newpage

\bibliography{references}
\end{document}